\documentclass[manuscript,sigconf,authorversion,nonacm]{acmart}
\usepackage{subcaption} 
\usepackage{csquotes}
\usepackage{enumitem}
\usepackage[most]{tcolorbox}
\usepackage{tabularx}
\usepackage{booktabs}

\AtBeginDocument{%
  }

\usepackage{csquotes}
\usepackage{enumitem}

\newtcolorbox{participantquote}[2][]{%
  enhanced,
  breakable,
  colback=gray!5,
  colframe=gray!40,
  arc=0mm,
  boxrule=0.5pt,
  left=4mm,
  right=2mm,
  top=2mm,
  bottom=2mm,
  fontupper=\itshape,
  before skip=8pt,
  after skip=8pt,
  title={\textbf{#2}},
  #1
}

\begin{document}

\title{Transformer-Based Interfaces for Mechanical Assembly Design: A Gear Train Case Study}

\author{Mohammadmehdi Ataei}
\affiliation{%
  \institution{Autodesk Research}
  \city{Toronto}
  \state{ON}
  \country{Canada}
}

\author{Hyunmin Cheong}
\affiliation{%
  \institution{Autodesk Research}
  \city{Toronto}
  \state{ON}
  \country{Canada}
}
\author{Jiwon Jun}
\affiliation{%
  \institution{Autodesk Research}
  \city{Belmont}
  \state{CA}
  \country{USA}
}

\author{Justin Matejka}
\affiliation{%
  \institution{Autodesk Research}
  \city{Toronto}
  \state{ON}
  \country{Canada}
}
\author{Alexander Tessier}
\affiliation{%
  \institution{Autodesk Research}
  \city{Toronto}
  \state{ON}
  \country{Canada}
}

\author{George Fitzmaurice}
\affiliation{%
  \institution{Autodesk Research}
  \city{Toronto}
  \state{ON}
  \country{Canada}
}
\renewcommand{\shortauthors}{Ataei et al.}

\begin{abstract}
Generative artificial intelligence (AI), particularly transformer-based models, presents new opportunities for automating and augmenting engineering design workflows. However, effectively integrating these models into interactive tools requires careful interface design that leverages their unique capabilities. This paper introduces a transformer model tailored for gear train assembly design, paired with two novel interaction modes: Explore and Copilot. Explore Mode uses probabilistic sampling to generate and evaluate diverse design alternatives, while Copilot Mode utilizes autoregressive prediction to support iterative, context-aware refinement. These modes emphasize key transformer properties (sequence-based generation and probabilistic exploration) to facilitate intuitive and efficient human-AI collaboration. Through a case study, we demonstrate how well-designed interfaces can enhance engineers' ability to balance automation with domain expertise. A user study shows that Explore Mode supports rapid exploration and problem redefinition, while Copilot Mode provides greater control and fosters deeper engagement. Our results suggest that hybrid workflows combining both modes can effectively support complex, creative engineering design processes.
\end{abstract}

\begin{teaserfigure}
  \includegraphics[width=\textwidth]{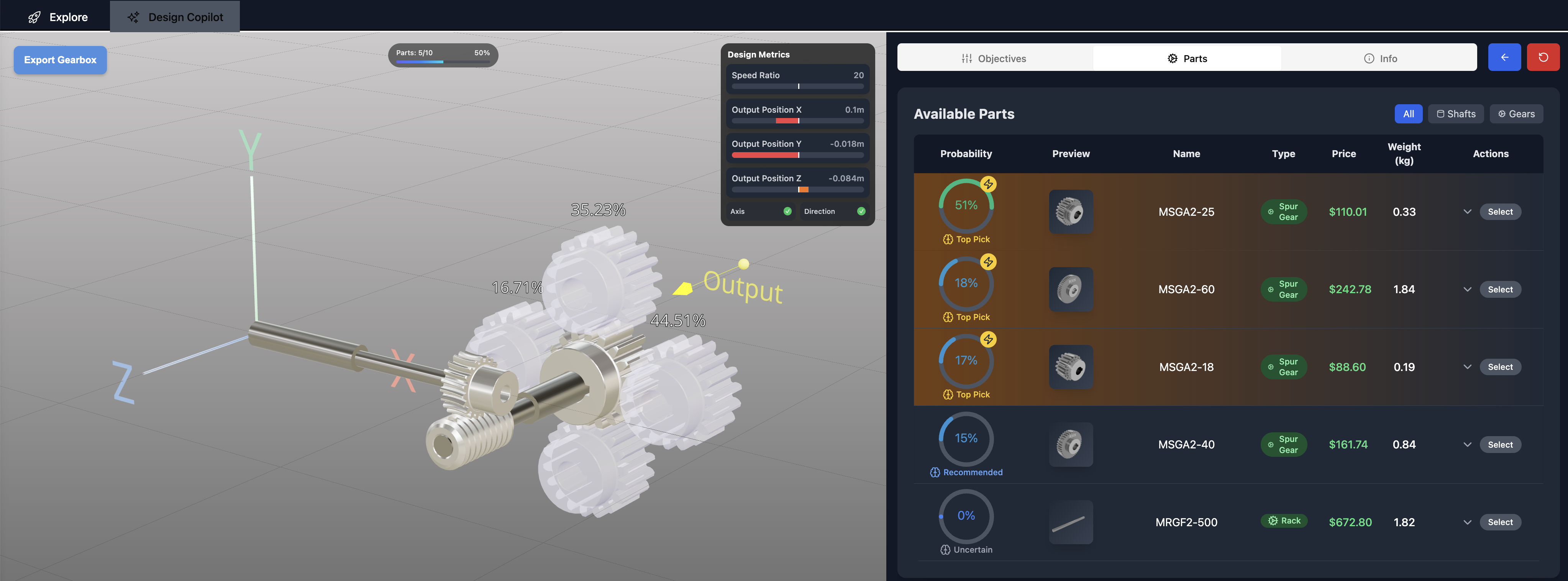}
    \caption{The Copilot Mode interface providing collaborative design refinement with GearFormer, featuring real-time 3D visualization (left), interactive design metrics panel (center top), and intelligent parts recommendation and selection interface (right).}
  \label{fig:teaser}
\end{teaserfigure}

\maketitle

\section{Introduction}
The integration of generative artificial intelligence (AI) into interactive design tools has the potential to transform how engineers and designers create complex mechanical assemblies. Traditional Computer-Aided Design (CAD) systems often require manual specification of every component, limiting the ability to rapidly explore diverse alternatives or efficiently navigate large, multidimensional design spaces. Transformer-based generative models have recently emerged as powerful tools for addressing these limitations. Their ability to generate structured, context-aware sequences makes them particularly effective for engineering applications involving constraints and interdependent components across domains such as mechanical assemblies, electronic circuit layouts, structural engineering, and materials discovery \cite{etesam2024deep, fu2023material, fang2025self}. These models excel in structured generation tasks due to their capacity to learn rich inter-component relationships and produce coherent, diverse outputs. However, integrating transformers into interactive user interfaces remains a challenge. Their unique characteristics (such as context-sensitive outputs, autoregressive generation, and probabilistic sampling) require new paradigms for human-AI interaction and interface design.

This paper explores how the fundamental properties of transformer-based models can inform the development of intuitive, efficient interfaces for assembly design. We focus on transformers’ strengths in autoregressive prediction and probability-driven sampling capabilities that enable real-time feedback, iterative refinement, and exploration of multiple valid alternatives. While other generative approaches, such as diffusion models \cite{yang2023diffusion}, may also support design workflows; however, their fundamentally different generation mechanisms necessitate distinct interaction strategies. In contrast, transformers’ step-by-step, interpretable output makes them well-suited for design tasks that require fine-grained, sequential decision-making.

To demonstrate these ideas, we present an interface built on top of GearFormer \cite{etesam2024deep}, a transformer-based model for gear train assembly design. We introduce two interaction modes tailored to its architecture: Explore Mode, which employs probabilistic sampling to generate diverse design alternatives, and Copilot Mode, which uses autoregressive generation to support context-aware, incremental design refinement. These modes illustrate how interface design can harness the specific strengths of transformers to support human creativity, guide complex decisions, and balance automation with expert judgment.

Through a case study and user evaluation, we show that interfaces grounded in model-specific capabilities can significantly improve the usability and impact of generative design tools. Our findings suggest that combining thoughtful interface design with the generative power of transformers enhances human-AI collaboration, enabling engineers to explore, iterate, and refine solutions within complex engineering workflows.

\section{Related Work}

\subsection{Human-AI Collaboration in Design}

There is tremendous value that AI can bring to augment and improve the design process. Yet, designers repeatedly face the challenge of adopting ever-evolving digital tools. \citet{palani_i_2022} identified practitioners' diverse motivations and concerns—ranging from "workflow integration" and "performance" to "emotional connection" with a tool—when deciding whether to adopt new creativity support technologies. In a more recent study, \citet{palani_evolving_2024} examined how generative AI is transforming creative workflows and perceived practitioner roles. They highlight tensions between automation and authorship, emphasizing that even with powerful AI systems, creative practitioners require human-centered interaction patterns for effective co-creation.

These findings resonate with those of \citet{shneiderman_human-centered_2022}: to maximize human benefit and maintain human control, AI-based systems must be designed with robust affordances that empower the user rather than replace them. Echoing this notion, \citet{mosqueira-rey_human---loop_2023} offer a broader review of human-in-the-loop machine learning, underscoring that the goal is to leverage AI for deeper collaboration, not to supplant human creativity. Our work directly addresses these concerns by designing specific interaction modes (Explore and Copilot) that explicitly leverage transformer model properties to facilitate such collaboration and maintain user control within the engineering design context.

\citet{bansal_beyond_2019} demonstrated that beyond system accuracy, people's mental models of a system's behavior strongly influence how well they calibrate trust and when they choose to override AI outputs. Similarly, \citet{gmeiner_exploring_2023} studied how engineering designers learn to work with AI-based manufacturing design tools, finding that interpretability, clarity in communicating design goals to the AI, and iterative exploration all remain central concerns. We aim to improve clarity and exploration by designing interfaces around the transformer's inherent sampling and sequential generation capabilities, making its behavior more predictable, clear, and useful for designers.

In a parallel line of research, \citet{lee_when_2025} proposed a framework linking design thinking stages to existing AI-based design support. While most current AI tools focus on generating solutions (e.g., conceptual geometry or final images), comparatively few provide robust assistance in discovering or defining the problem domain. Reflecting on these insights, \citet{shi_understanding_2023} offered a systematic literature review of collaborative AI in design, highlighting the need to clarify roles (e.g., AI assisting designers vs. designers assisting AI), as well as new methods for effectively merging AI-based generation with domain-specific human insight. Our Explore Mode seeks to aid in problem definition through iterative exploration, while Copilot Mode focuses on merging AI generation with human insight during solution refinement, addressing different stages of the design process identified by these prior works.

\subsection{Human-in-the-Loop Design Exploration}

Computational systems can amplify the user's ability to explore the design space by generating multiple alternative solutions. Much of the work in this domain aims to assist the user in exploring the vast set of designs produced by such systems. Early examples include \cite{lee_designing_2010}, who introduced interactive example galleries to help non-expert users discover, adapt, and integrate existing design ideas into their own work. Similarly, \citet{koyama_crowd-powered_2014} pioneered a crowd-powered parameter analysis method, in which crowd feedback on parameter settings aids users in exploring high-dimensional design spaces more effectively. The Dream Lens system by \citet{matejka_dream_2018} targeted the exploration of generative design datasets at scale, enabling the sorting and visualization of thousands of automatically generated design solutions. In the robotic design domain, \citet{desai_geppetto_2019} showcased that a generative design process can be augmented with crowd-sourced emotional perception of robot motions, therefore supporting the semantic design of expressive robot behaviors that are subjective by nature and can be difficult to be achieved with computational methods alone. Later, \citet{koyama_sequential_2020} extended these concepts with the Sequential Gallery, an iterative framework that incorporates Bayesian Optimization to reduce search complexity; users can explore two-dimensional "slices" of large design spaces in a more manageable way. Our Explore Mode interface builds upon these design exploration concepts, specifically leveraging the probabilistic sampling unique to transformer models to generate diverse gear train alternatives and employing visualizations like Pareto fronts to help users navigate the resulting design space effectively.

\subsection{Interactive Generative Design Systems}

Developing user-driven generative design systems is another important  research area. DreamSketch by \citet{kazi2017dreamsketch} was a pioneering work at the intersection of sketching and 3D generative design that aims to capture the essence of the early-stage design stage typically done with sketching and the detailed design stage where precise computations are used to find optimal 3D geometries. \citet{chen_forte_2018} introduced Forte, a sketch-based, real-time topology optimization system that allows designers to visualize structural changes on the fly. Meanwhile, \citet{khan_genyacht_2019} proposed GenYacht, a hybrid system combining generative search and iterative user selection for optimizing yacht hull designs. Each of these works emphasizes the importance of humans actively participating in the generative design process that marries the designer's aesthetic or contextual requirements with the computational power of generative design systems. Inspired by these systems, our Copilot Mode provides an interactive environment, but rather than relying on sketches or global search, it harnesses the transformer's autoregressive capabilities to offer context-aware, step-by-step component recommendations, facilitating a tight interactive loop between the designer and the AI during assembly creation.

\subsection{Deep Generative AI for Engineering Design}

Recently, applications of generative AI for engineering design have gained significant traction, and multiple deep generative models have been developed to solve various design problems \cite{regenwetter2022deep}. These problems include structural design \cite{giannone2023aligning, picard2024generative, raina2023learning}, aerodynamic airfoil design \cite{chen2019synthesizing}, kinematic linkage design \cite{lee2023deep}, and terrain-optimized robotic design \cite{zhao2020robogrammar}.

\section{Designing User Interfaces for Transformer-Based Generative Models}

Developing successful user interfaces for generative AI systems hinges on thoroughly understanding the underlying models and their unique characteristics. Different generative approaches, such as diffusion models \cite{yang2023diffusion} that iteratively refine noise into coherent outputs, or Generative Adversarial Networks (GANs) \cite{goodfellow2020generative} that learn through competition, possess distinct internal mechanisms. Consequently, interaction strategies and interface designs must capitalize on the specific properties of the generative approach in use to be effective. For example, an interface designed for a diffusion model might emphasize iterative refinement controls, while one for a GAN might focus on exploring the latent space.

\subsection{Understanding GearFormer}

GearFormer is built on the transformer architecture \cite{attentionisallyouneed}, originally designed for natural language processing. Transformers employ a self-attention mechanism that dynamically weights different parts of an input sequence in context. Unlike recurrent neural networks (RNNs) \cite{yu2019review}, which process sequences incrementally, transformers can attend to every position in the sequence simultaneously, enabling the capture nuanced patterns and dependencies.

Transformers can take different forms (encoder-only, decoder-only, or encoder-decoder). GearFormer \cite{etesam2024deep} adopts an encoder-decoder structure where the encoder processes the entire input sequence (e.g., design requirements) at once, and the decoder generates output tokens sequentially.

First, the encoder captures the input context, such as design constraints and objectives. Then, at each step of generation, the decoder relies on the encoded information and previously generated tokens to propose a distribution over all possible next tokens. Rather than outputting a single next token, GearFormer outputs a \emph{probability distribution} across its vocabulary, which encodes the likelihood of various design elements to be included next. For example, the model might assign 65\% probability to one gear type, 25\% to another, and spread the remaining 10\% among other possibilities.

GearFormer's vocabulary is composed of two main token categories. The first category, \emph{component selection tokens}, determines which parts (e.g., spur gear, bevel gear, shaft) should next appear in the design. The second category, \emph{positioning tokens}, specifies placement and orientation details for each component. By separating these lexicons, the model can simultaneously handle constituent decisions (which components to include) and spatial decisions (where to arrange them). In general, after a component token for a particular gear type is decided, the model produces positioning tokens that establish its location in the assembly.

Similar to language, a valid sequence of tokens produced by the model must follow a specific grammar. For example, the grammar enforces constraints on gear meshing, compatible gear types (e.g., spur-to-spur, bevel-to-bevel), shaft connections, and interference-free assemblies. During training, the model is exposed to data that always conforms to this grammar, so it gains the capability to produce sequences that always conform to the grammar.

As mentioned earlier, transformers output a probability distribution for the next token choice. Based on this feature, it can generate a sequence in two ways: (1) greedy generation, in which the model always picks the most likely token and yields a consistent, if potentially less creative, sequence; and (2) stochastic generation, where each token is sampled from the probability distribution with varying degrees of randomness, promoting the diversity and exploration at the expense of straying from the expected sequence given the input. The randomness is mainly controlled by a parameter called \emph{temperature}: higher temperatures yield more varied outcomes, while lower temperatures produce more predictable, expected sequences.

\begin{figure*}
    \centering
    \includegraphics[width=1.0\linewidth]{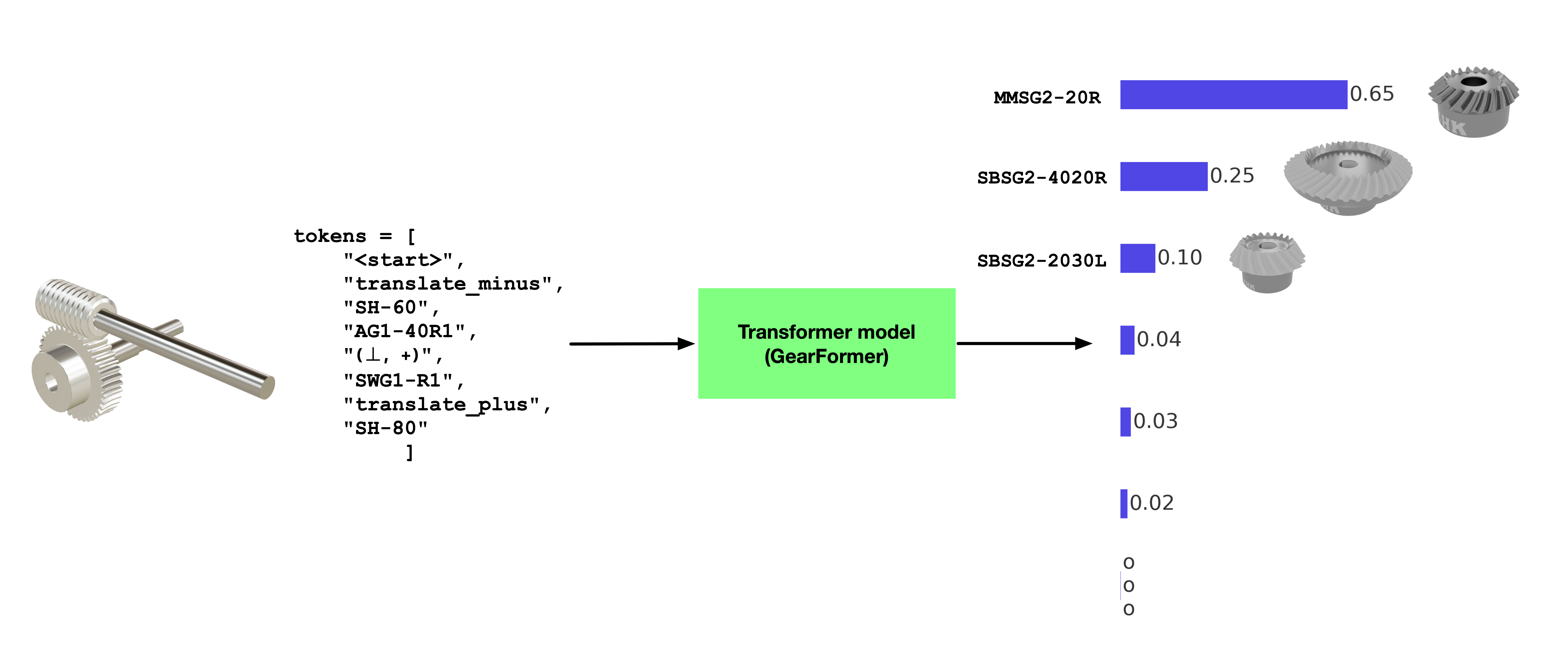}
    \caption{An illustration of GearFormer's inference process. A tokenized representation of the partially assembled gear mechanism is passed into a transformer model, which predicts the probability distribution over the next possible components. The next component is selected either as the most probable or by sampling from the distribution. The top predictions and their corresponding probabilities are shown on the right.}
    \label{fig:token-probs}
\end{figure*}

As each token is added, it forms part of the growing context for subsequent predictions, establishing a continuous feedback loop. This iterative process continues until the model produces a full design or reaches a stopping criterion. Figure~\ref{fig:token-probs} illustrates this process. A sequence of tokens encoding spatial and component relationships is fed into a transformer model, GearFormer. The model then outputs a probability distribution over possible gear components. The next component is selected either as the most probable one or by using a sampling method, and the process repeats to gradually construct the full assembly.

\subsection{Key Properties of GearFormer for Interface Design}

The following sections detail how the inherent properties of GearFormer inform the user interface design and highlight key considerations when creating interfaces for transformer-based generative models.

\subsubsection{Sampling and Exploration}

\begin{figure*}[ht]
  \centering
  \includegraphics[width=\textwidth]{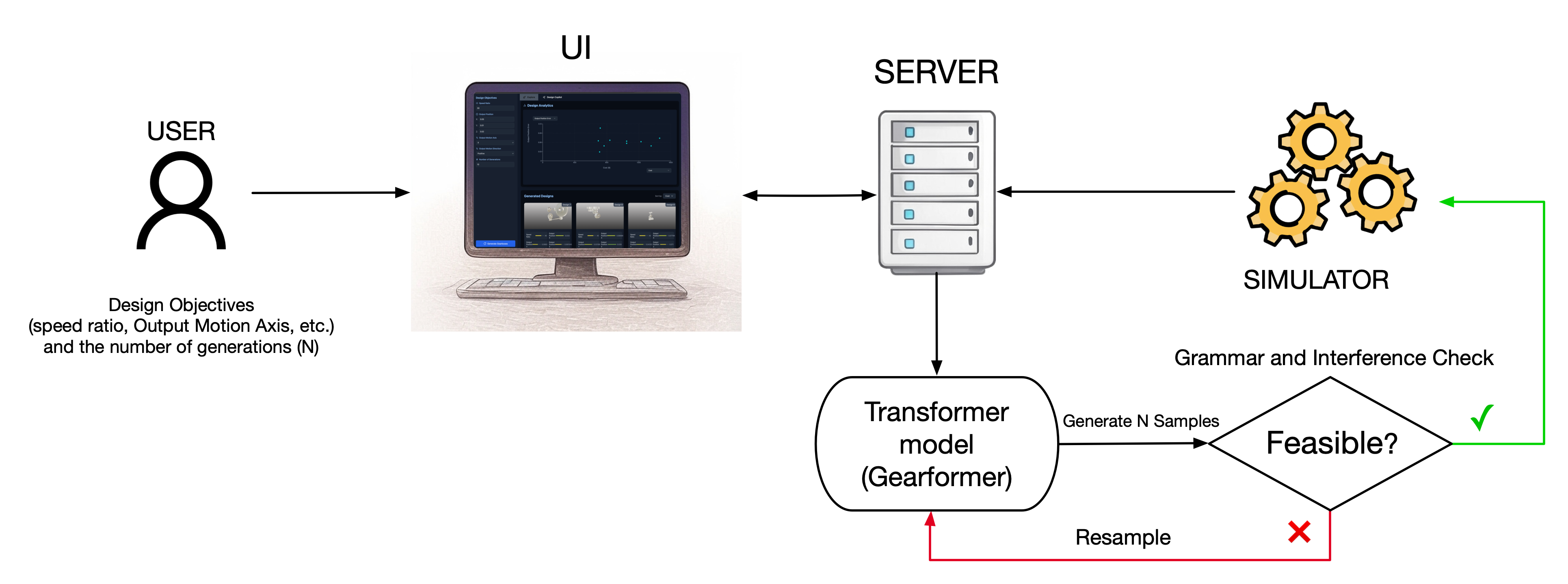}
  \caption{Schematic overview of Explore Mode's sampling-based workflow. Users define design objectives and constraints, prompting GearFormer to generate multiple candidate assemblies, which are validated and then presented for rapid comparison.}
  \label{fig:explore-mode-diagram}
\end{figure*}

To demonstrate the transformer's capacity to generate diverse designs, we created Explore Mode, depicted in Figure~\ref{fig:explore-mode-diagram}. The user starts by defining high-level design objectives such as desired speed ratio, output placement, and the number of alternative designs to generate. GearFormer then produces multiple solutions via controlled temperature sampling, capturing a range of possible trade-offs among competing objectives. For example, the perfect adherence to all constraints may be impossible, so each design represents a distinct compromise.

Like any generative model, GearFormer can produce incorrect outputs, though such cases are rare. Generating designs that violate grammatical (e.g., connecting incompatible components) or physical feasibility (e.g., part collisions) is undesirable and may lead to confusion for the user. To address this, Explore Mode integrates a simulator that validates each assembly. Any infeasible proposals are filtered out, prompting resampling until the requested number of valid alternatives is reached. This approach gives users the freedom to explore unconventional possibilities without being overwhelmed by infeasible concepts, creating a reliable yet creative design exploration experience.

\subsubsection{Autoregressive Recommendation}

\begin{figure*}[ht]
  \centering
  \includegraphics[width=\textwidth]{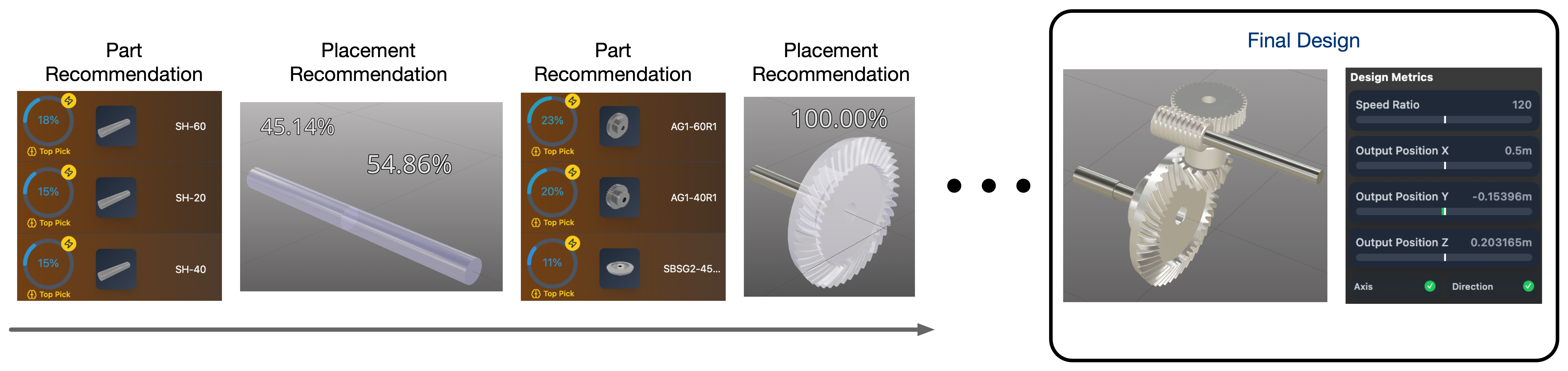}
  \caption{Overview of Copilot Mode's iterative design workflow. At each step, GearFormer provides ranked part and placement recommendations based on its autoregressive model. Designers select components and positions with confidence overlays, progressively building the assembly. Real-time feedback on feasibility and alignment with design objectives—such as speed ratio and output position—is shown alongside the evolving 3D model. The process continues until a complete, validated gear train is assembled.}
  \label{fig:copilot-mode-diagram}
\end{figure*}

As mentioned earlier, transformers make the prediction of each token autoregressively (i.e., the next token prediction depends on the prior token choices). One could then leverage this feature to employ transformers as advanced recommender systems. Given any partial design, GearFormer predicts the probability distribution of next tokens for both component selection and spatial positioning. The choice for the next token could then be recommended to the user based on the probability distribution. For instance, if a user chooses a specific gear and its position, GearFormer can propose subsequent components with the highest probabilities that would fulfill the design requirements. Each choice made by the user would then lead the model to recalculate the probability distribution of the next tokens, and hence the next recommendations of components and their placements.


To illustrate this recommender functionality, we developed the Copilot Mode, shown in Figures~\ref{fig:teaser}~and~\ref{fig:copilot-mode-diagram}. The user specifies main design goals such as speed ratios and output positions, and the model suggests the next component and placement options ranked by their probabilities. The designer can choose from these recommendations one decision at a time, and GearFormer recalculates and suggests next compatible components and placements.

This human-in-the-loop process ensures that designers retain control over the final design, using AI suggestions as an adaptive guide rather than a strict prescription for the whole assembly design. The resulting interplay blends algorithmic insights with human creativity and domain expertise, yielding solutions that meet engineering constraints while honoring the designer's intent.

\section{User Interface Design for GearFormer}

This section goes into the details of: Explore Mode and Copilot Mode. Explore Mode focuses on generating and comparing diverse design alternatives through transformer sampling, while Copilot Mode centers on incremental, user-guided design completion enabled by GearFormer's autoregressive "autocomplete" properties. Both interfaces showcase the model's capacity for handling structured mechanical assemblies, integrating domain-specific constraints in real-time.

\subsection{Explore Mode Interface}

\begin{figure*}[ht]
    \centering
    \includegraphics[width=\textwidth]{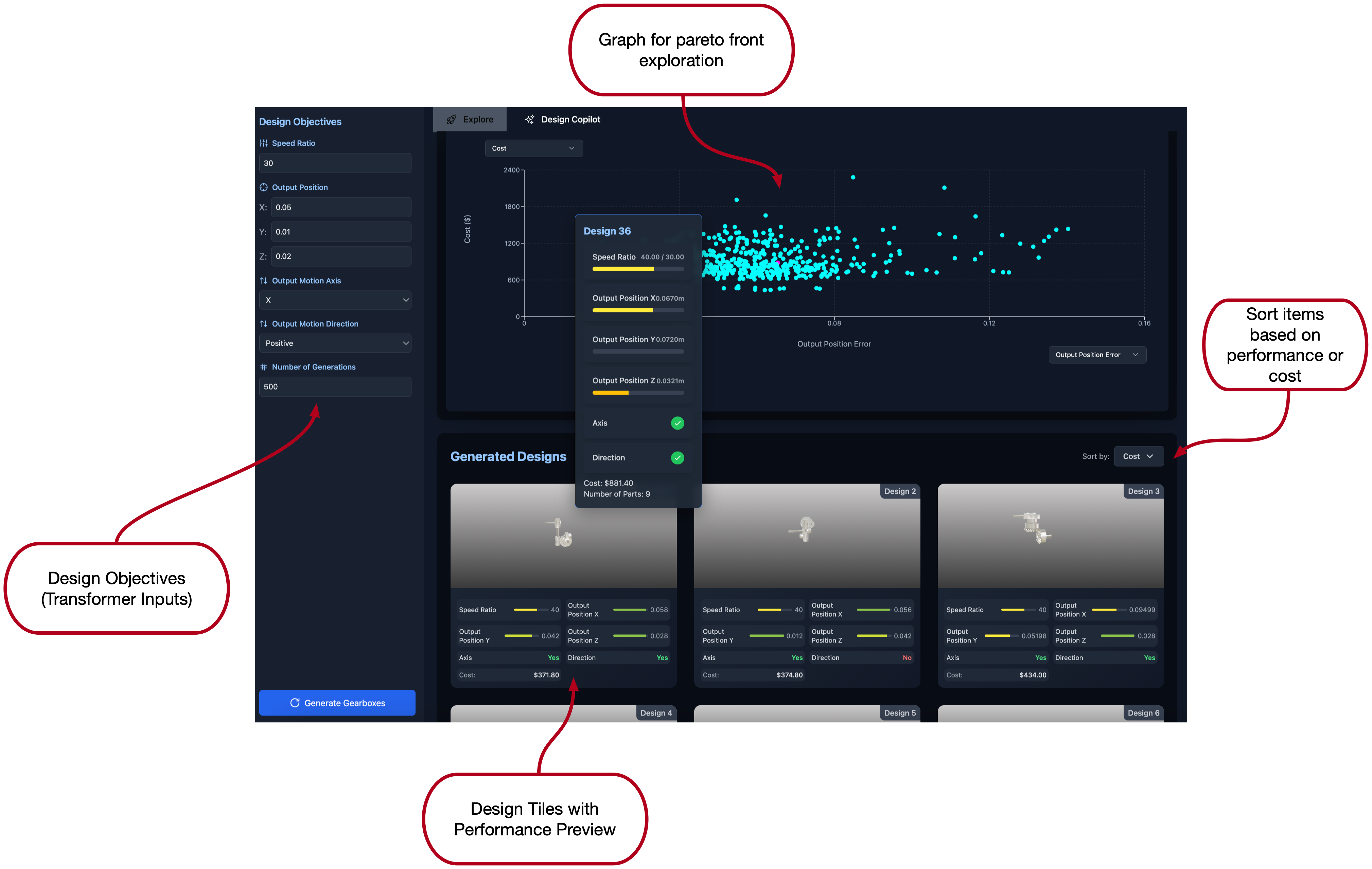}
    \caption{Explore Mode interface, illustrating the Design Objectives panel for specifying transformer inputs (left), an interactive Pareto front graph for visualizing trade-offs (top center), Generated Design Tiles for rapid performance comparisons (bottom), and sorting controls for ranking designs by cost or performance (right).}
    \label{fig:explore-mode-interface}
\end{figure*}

Explore Mode (Figures~\ref{fig:explore-mode-diagram}~and~\ref{fig:explore-mode-interface}) is designed to exploit GearFormer's ability to sample diverse design alternatives from its learned probability distributions. This sampling approach enables users to explore a wide variety of gear train layouts—potentially uncovering innovative solutions that might not emerge through purely deterministic methods.

\paragraph{Design Objectives Panel}
Located on the left side of the interface, the Design Objectives panel lets users enter key transformer inputs, including the target speed ratio, output position coordinates, the output motion axis, and direction. Users may also specify the number of design generations (e.g., 500) they wish to explore. Since transformer models require carefully structured token inputs, clarity in these high-level parameters ensures that the model's resulting output space remains aligned with the user's engineering requirements.

\paragraph{Generated Design Tiles and Performance Preview}
At the bottom of the interface, Explore Mode displays multiple design alternatives as distinct "tiles." Each tile presents a concise performance summary encompassing:
\begin{itemize}
    \item \textbf{Speed Ratio:} How close the generated design is to meeting the specified ratio.
    \item \textbf{Output Positions (X, Y, Z):} Summaries of how precisely the design matches target coordinates.
    \item \textbf{Axis and Direction Checks:} Indicators signaling whether the design's motion axis and direction align with user preferences.
    \item \textbf{Cost Estimates:} A quick reference for financial feasibility.
\end{itemize}
By presenting numerous alternatives simultaneously, Explore Mode capitalizes on GearFormer's sampling capability: rather than converging on a single deterministic solution, it offers a breadth of designs so that engineers can quickly gauge potential trade-offs.

\paragraph{Interactive Pareto Front Graph}
Shown in the top center of Figure~\ref{fig:explore-mode-interface}, a dynamic scatter plot provides high-level visualization of trade-offs between performance metrics (e.g., output position error) and cost. Each plotted point represents an individual design generated by GearFormer. Users can hover over or click data points to view key metrics in real time, facilitating immediate identification of promising or outlier solutions. The axes of this plot can be reconfigured to compare any pair of metrics (e.g., speed ratio vs. position error, or cost vs. weight), allowing users to explore different dimensions of the design space. This graph clarifies how certain performance gains might come at the expense of higher costs, or vice versa, thus highlighting the multi-objective nature of gear train design.

\paragraph{Sorting and Filtering Controls}
To the right of the interface, sorting and filtering tools allow users to prioritize designs based on specific objectives (e.g., lowest cost or smallest position error). Given the potentially large number of generated designs, these tools help designers effectively navigate and refine the solution space, increasing the efficiency of the decision-making process.

\subsubsection{Detailed Design Inspection in Explore Mode}

\begin{figure*}[ht]
    \centering
    \includegraphics[width=\textwidth]{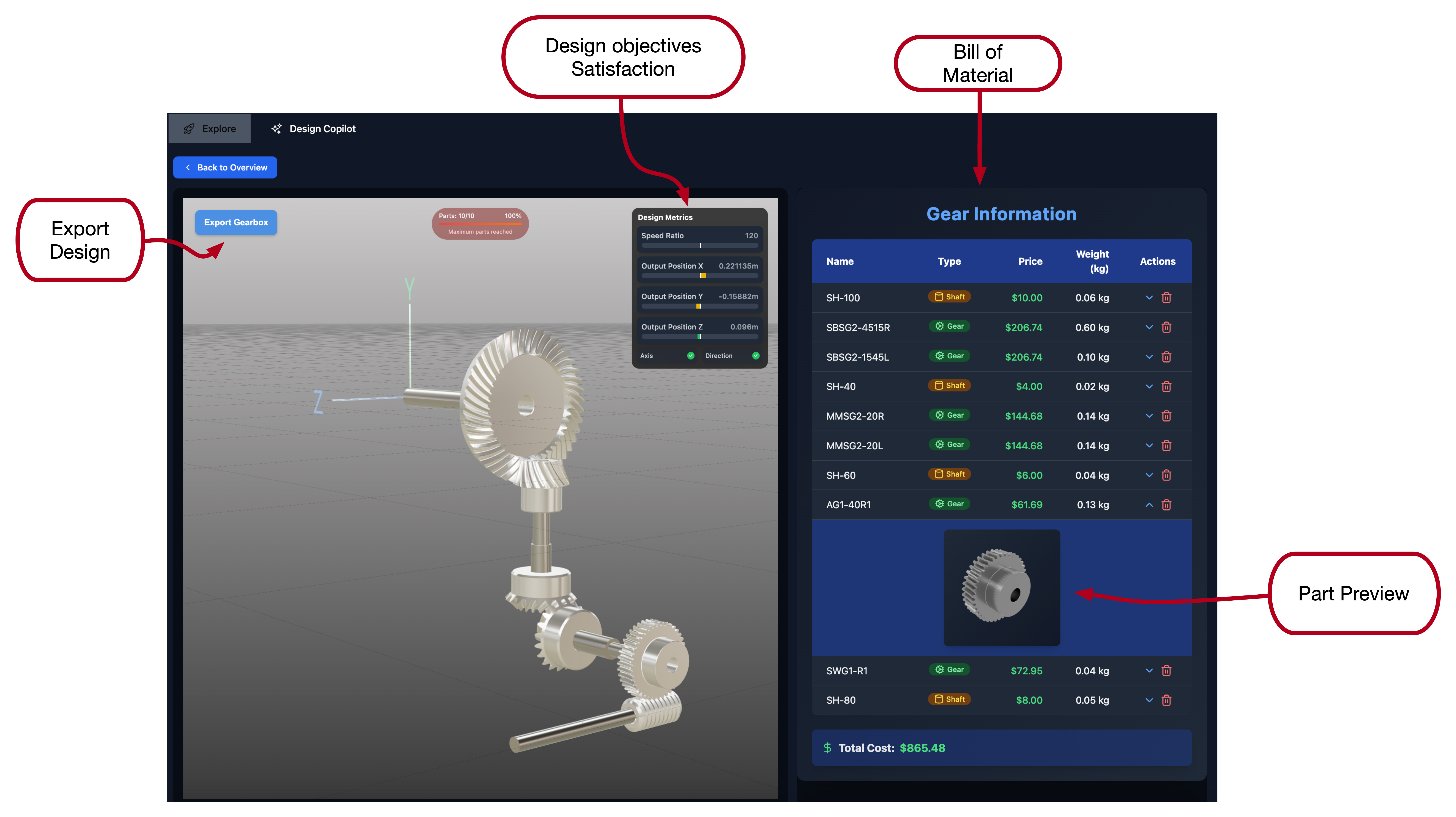}
    \caption{Detailed inspection interface for Explore Mode. A selected design is rendered in an interactive 3D view (left), displaying each gear and shaft. A Bill of Materials (right) lists individual components by name, cost, and weight, while a metrics panel (top center) reports how closely the design meets the user's speed ratio and position targets.}
    \label{fig:design-detail-interface}
\end{figure*}

While Explore Mode initially provides an overview of many design variations, each tile can be further examined by entering a dedicated inspection view. Figure~\ref{fig:design-detail-interface} showcases this deeper analysis environment.

\paragraph{Interactive 3D Visualization}
A real-time, manipulable 3D model of the current design is displayed, allowing users to rotate, zoom, and examine gear alignments from multiple angles. This close-up view is vital for spotting spatial interferences or misalignments that may not be evident in the aggregated performance previews.

\paragraph{Design Metrics Panel}
Displayed adjacent to the 3D model, this panel enumerates crucial objective satisfaction indicators. Engineers can confirm whether the realized speed ratio is within acceptable bounds, check X/Y/Z alignment errors, and confirm that the motion axis and direction match the specified constraints. These instantly updated metrics help designers quickly decide if further refinements are needed.

\paragraph{Bill of Materials and Part Previews}
On the right, a Bill of Materials details every gear, shaft, and additional part in the design. Users can see each part's name, type, price, and weight at a glance. Selecting a specific component opens a mini-preview of the part, reinforcing clarity about its geometry and orientation. This granular breakdown assists in evaluating both technical feasibility and cost implications.

\paragraph{Export Functionality}
Finally, engineers can export the fully assembled design (including geometry, metadata, and component list) for downstream simulation, documentation, or manufacturing workflows. This immediate export feature supports integration of generative outputs into standard engineering pipelines.

\section{Copilot Mode Interface}

Copilot Mode, shown in Figure~\ref{fig:teaser}, offers an iterative, user-in-the-loop environment that capitalizes on GearFormer's autoregressive properties. Whereas Explore Mode focuses on presenting diverse, pre-generated solutions, Copilot Mode enables continuous collaboration between human designers and the underlying transformer model, akin to an advanced "autocomplete" for mechanical assemblies.

Copilot Mode emphasizes active collaboration, aligning well with the transformer's strength in token-by-token autocomplete. By systematically narrowing design uncertainty at each step—leveraging immediate model feedback, engineers can guide GearFormer to viable solutions that might not arise from unguided sampling alone. The interface's fluid, interactive structure also keeps the human designer firmly in control: final decisions rest with the user, ensuring that the process remains grounded in domain expertise and practical considerations, while still benefiting from AI-driven insights.

\subsection{Key Features of Copilot Mode}

\paragraph{Real-Time 3D Visualization Workspace}
At the interface's center is a large, interactive 3D workspace (Figure~\ref{fig:teaser}). Users can directly observe and manipulate the evolving gear train design. Whenever a user selects or places a gear, GearFormer recalculates subsequent component choices based on its learned domain-specific constraints, immediately updating the design's 3D representation.

\paragraph{Intelligent Parts Recommendation and Selection}
One of the most powerful features in Copilot Mode is the parts recommendation panel. Building upon GearFormer's probabilistic token outputs, the interface ranks and categorizes suggested components—gears, shafts, bevel gears, and more—according to their likelihood of fitting well in the current design context. Users can:
\begin{itemize}
    \item Browse detailed specs (price, weight, geometry).
    \item Accept a recommended part (prompting the model to place it accordingly).
    \item Override the recommendation by selecting a different component, forcing the model to adapt subsequent tokens.
\end{itemize}
This mixed-initiative approach leverages both human intuition and the transformer's learned patterns, ultimately producing configurations that reflect realistic engineering constraints, user preferences, and design creativity.

\paragraph{Placement Confidence Overlays}

\begin{figure}[ht]
  \centering
  \includegraphics[width=\linewidth]{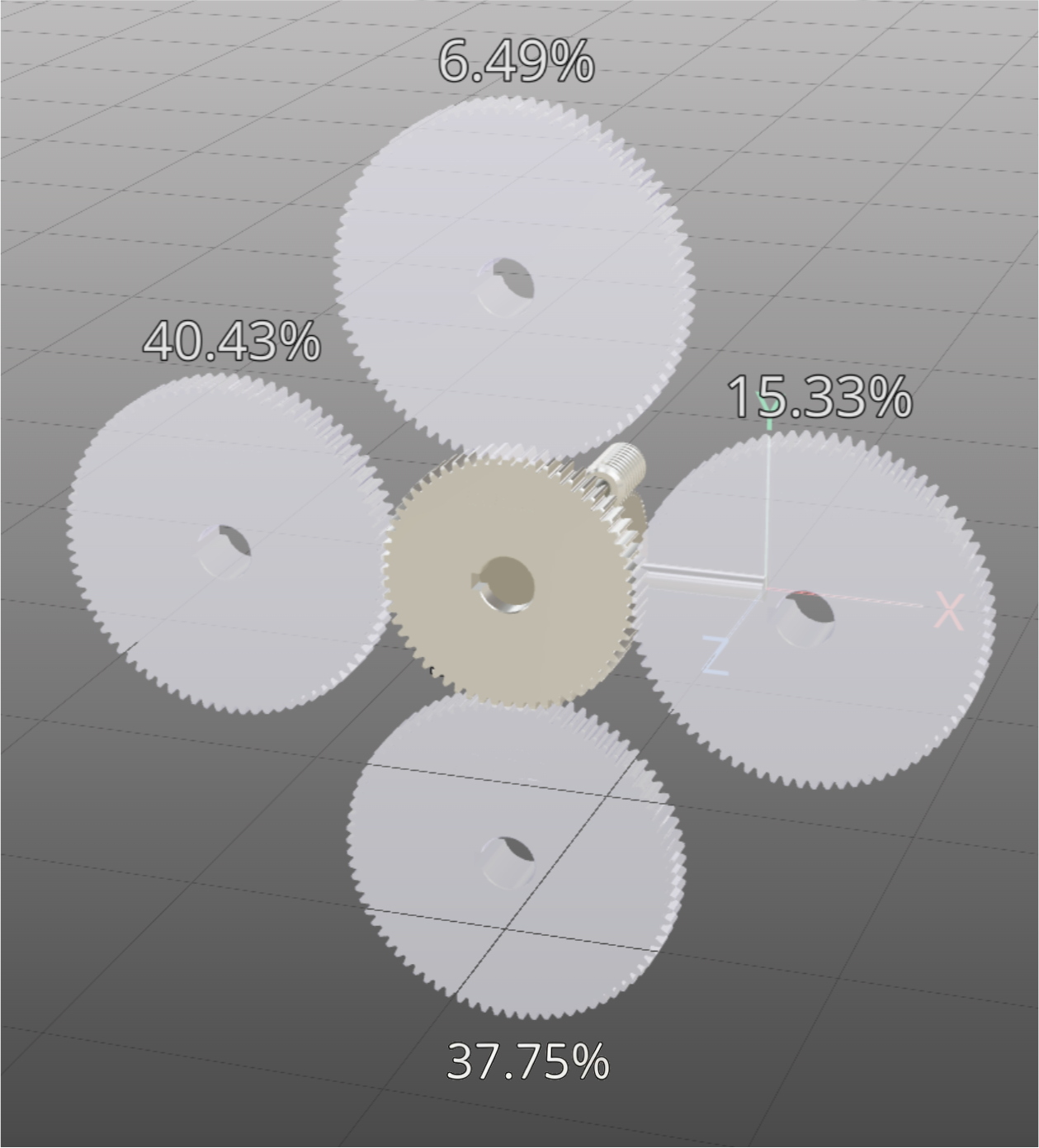}
  \caption{Example of a gear placement with a confidence score overlay in Copilot Mode.}
  \label{fig:gear-p}
\end{figure}

As shown in Figure~\ref{fig:gear-p}, when a user places a gear, the interface overlays a confidence score on it, based on the transformer's confidence in the placement. Possible gear placements are indicated by transparent gears, each accompanied by a percentage score displayed nearby. When the user clicks to select a gear, the chosen transparent gear becomes solid, and all other gears are removed.

\paragraph{Design Metrics Panel}
A dedicated panel continuously monitors and displays performance metrics, including speed ratio accuracy, positional coordinates, and compliance with user-specified axis and direction requirements. These metrics are updated as soon as a user alters any component or the model generates a new layout, allowing engineers to track whether each iterative change moves the design closer to or further from desired objectives.

\begin{figure}[ht]
  \centering
  \includegraphics[width=\columnwidth]{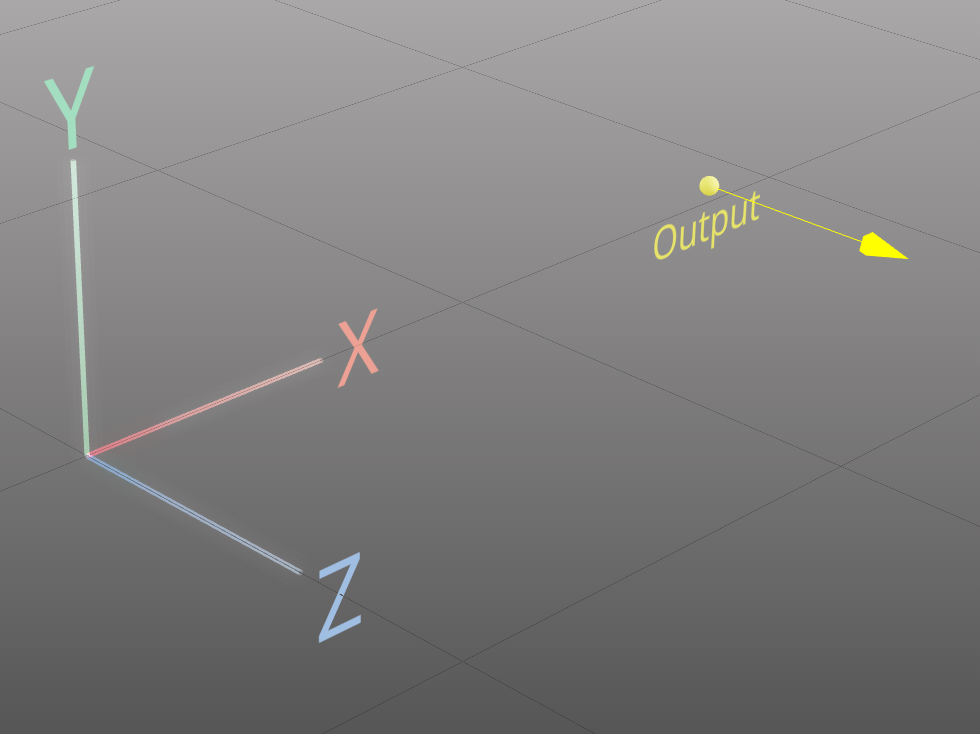}
  \caption{The yellow ``Output'' arrow visualizes the target position and orientation for the gear train output, serving as a spatial reference point for designers. This directional indicator helps users align their design with the required motion axis and position specified in the design objectives.}
  \label{fig:output-dir}
\end{figure}

\paragraph{Output Direction Visualization}
A prominent yellow arrow labeled ``Output'' (visible in Figures~\ref{fig:teaser}~and~\ref{fig:output-dir}) indicates the target position and orientation for the gear train's output. This visual cue helps designers understand both the spatial location (X, Y, Z coordinates) and the required motion axis/direction specified in the design objectives. This goes in hand with the Design Metrics Panel interface that highlights position errors in real-time, allowing designers to see how closely their current design approaches the target output position and orientation.

\paragraph{Automatic Updates and Feedback}
Each accepted or rejected recommendation triggers an immediate cascade of recalculations. GearFormer updates the design representation and confidence estimates for the remaining components, enabling a smooth, continual design flow.

\paragraph{Context Limits and Maximum Part Count}

\begin{figure}[ht]
  \centering
  \includegraphics[width=0.6\linewidth]{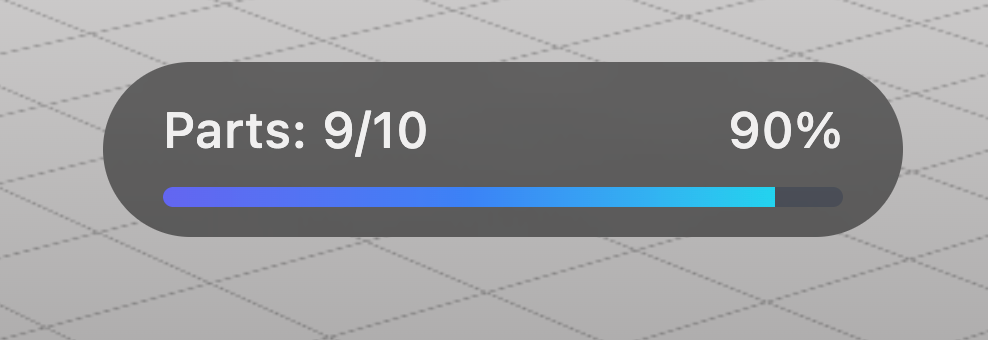}
  \caption{Example of a part usage indicator in the interface (here shown as 9 parts used out of a maximum 10). Exceeding the maximum limit can cause a transformer to produce inconsistent tokens due to context constraints.}
  \label{fig:max-parts}
\end{figure}

A small progress or capacity indicator (similar to Figure~\ref{fig:max-parts}) also alerts users when they approach the part limit, minimizing the risk of exceeding the transformer's context length. GearFormer, being built upon a transformer model, processes input tokens within a fixed context window, determined during model training. If a design grows too large (exceeding the token capacity required to represent each part's position, type, and relationships) the model may struggle to maintain coherence across all constraints. As illustrated in Figure~\ref{fig:max-parts}, the interface explicitly enforces a maximum part limit. If users approach or surpass this threshold, an on-screen indicator warns them that additional components may be ignored or produce invalid tokens. This mechanism prevents the model from generating impractical designs by overextending its context capacity, ensuring that all recommended solutions remain grounded in physically consistent gear assemblies.

\paragraph{Exporting the Completed Gearbox}
Once a satisfactory layout is achieved, a single click allows designers to export the finished gear train model as a CAD file.


\section{Implementation Details and Performance}

Our implementation adopts a client-server architecture, separating computationally intensive model operations from the user-facing interface components.

The backend is developed using FastAPI, a modern and high-performance Python web framework, enabling efficient handling of multiple concurrent client requests in a non-blocking manner.

The backend hosts the GearFormer model and exposes REST API endpoints for design generation, validation, and evaluation. In Copilot Mode, additional endpoints support incremental design updates and real-time component recommendations.

The frontend is built with React and Next.js. For 3D visualization of gear assemblies, we utilize Three.js, a JavaScript 3D library leveraging WebGL for hardware-accelerated browser-based rendering.

The GearFormer backend operates on an AWS EC2 \texttt{g4dn.xlarge} instance equipped with an NVIDIA Tesla T4 GPU.

In Explore Mode, the backend concurrently generates multiple complete designs in parallel batches, processing each request in approximately \texttt{760ms}, including model inference, design validation the simulator, and server-client communication.

In Copilot Mode, designed specifically for real-time interaction, end-to-end response time of approximately \texttt{100ms} (encompassing model inference and server-client communication) is achieved by retaining the model state in memory during next token generation, significantly improving responsiveness by avoiding redundant computations.

\section{User Study}

We conducted a user study to gain qualitative observations about the general design workflows afforded with a generative AI model and compare the benefits and limitations of the two UI modes developed.

\subsection{Participants}

Ten participants were recruited for the study (two female, eight male). Two participants (P1 and P2) were experienced gear train designers with more than 10 years of industry CAD experience. The other eight participants (P3-P10) had varying degrees of engineering design and CAD experience ranging from 3 to 10 years. Each participant was reimbursed \$75 USD or equivalent value for their time.

\subsection{Study Design}

We used a within-subject design by asking each participant to complete a design task with both Explore and Copilot modes in sequence. The order of the first mode used was counterbalanced.

\begin{figure}[h!]
\centering
\includegraphics[width=\columnwidth]{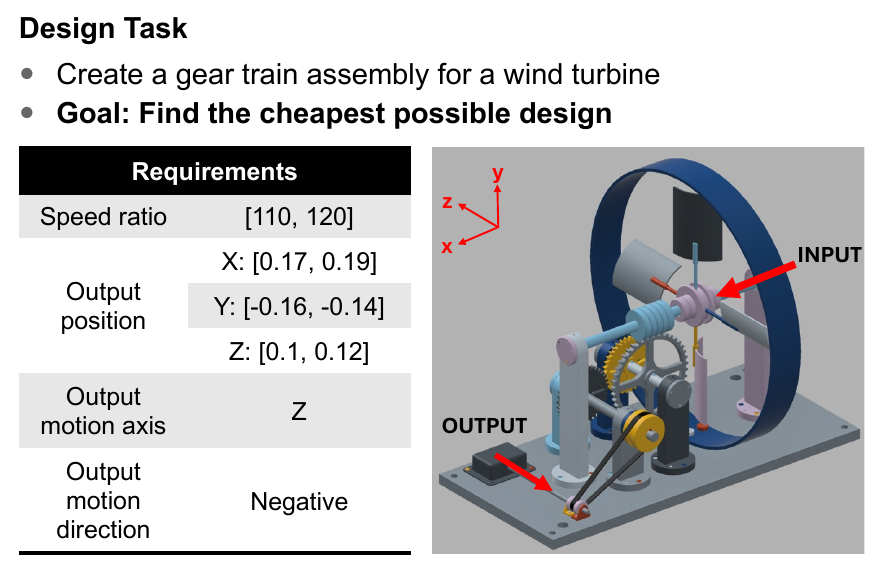}
  \caption{Design task given to the participants.}
  \label{fig:task}
\end{figure}

The design task involved creating a new gear train design for a wind turbine model, shown in Figure~\ref{fig:task}. Their goal was to find the cheapest possible design that met the requirements given. For the speed ratio and output position requirements, they were told any number that falls between the specified ranges would suffice. The requirement metrics are chosen intentionally such that meeting all the requirements is not trivial and the participant would require multiple interactions with the UI to find a feasible solution.

\subsection{Procedure}

The study began with a general introduction, questions about participants' CAD and gear train design experience, and the introduction of GearFormer. Then, we performed a walk-through of the UI in the first mode, followed by presenting them the design task. Ten minutes were given to complete the design task using the first mode. We then asked interview questions specifically about the first mode that they tried. Afterwards, we repeated the walk-through for the second mode and gave the participants 10 minutes to complete the design task with the second mode. We again asked questions specifically about the second mode, and ended the study with additional questions comparing the two modes. The whole study lasted an hour.

\subsection{Interview Questions}

For each mode, we asked the following interview questions. The first two questions were intended to gather additional insights that may not have been observed while they are using the UI, while the last question was asked to gauge their trust in the AI tool.
\begin{itemize}
\item Q1: What did you find useful about the [Explore/Copilot] mode?
\item Q2: What did you find challenging about the [Explore/Copilot] mode?
\item Q3: How confident do you feel about the gear train designs you generated using the [Explore/Copilot] mode? Please rate on a scale from 1 to 5, where 1 is not confident at all and 5 is very confident.
\end{itemize}

In addition, the following interview questions were asked to compare the two modes:
\begin{itemize}
\item Q4: Between the two modes, which mode did you find easier to use?
\item Q5: Between the two modes, which mode helped you gain more knowledge about gear train designs?
\item Q6: Let's imagine that you're a mechanical designer creating gear train designs on a daily basis. If you had to use one mode for your work, which one would you choose?
\end{itemize}

While participants were encouraged to pick only one mode as the answer to these questions, several participants (3) gave more nuanced answers, e.g., ``If I was an experienced gear train designer, the Copilot Mode, but if I was a beginner, the Explore Mode''. In such a case, we counted half a point for each mode as the answer.

\subsection{Quantitative Findings}

\begin{figure}[h!]
    \centering
    \begin{subfigure}[b]{0.48\columnwidth}
        \centering
        \includegraphics[width=\linewidth]{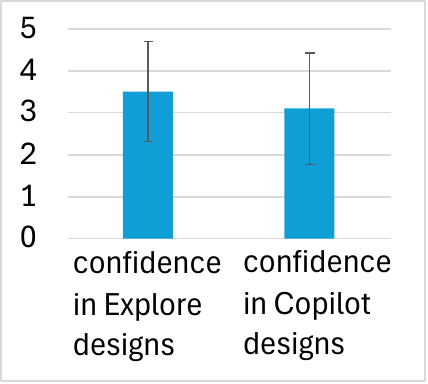}
        \caption{}
        \label{fig:conf}
    \end{subfigure}
    \hfill
    \begin{subfigure}[b]{0.47\columnwidth}
        \centering
        \includegraphics[width=\linewidth]{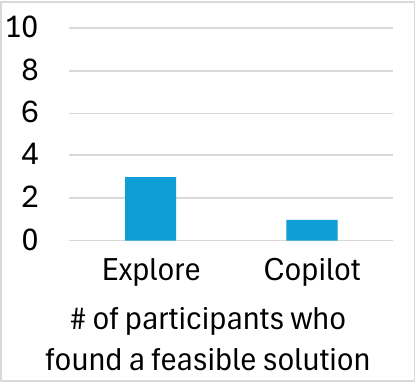}
        \caption{}
        \label{fig:success}
    \end{subfigure}
    \caption{For each respective mode, (a) average participant ratings [0-5] on their confidence in the designs generated and (b) number of participants who found a feasible solution that met all requirement metrics.}
\end{figure}

In response to Q3, Figure~\ref{fig:conf} shows that participants tend to give a higher confidence rating to the designs found with the Explore Mode. However, the difference was not statistically significant (M = 3.5, SD = 0.85 for Explore and M = 3.1, SD = 1.20 for Copilot, t(18) = 0.86, p = .41).

\begin{figure}[h!]
\includegraphics[width=\columnwidth]{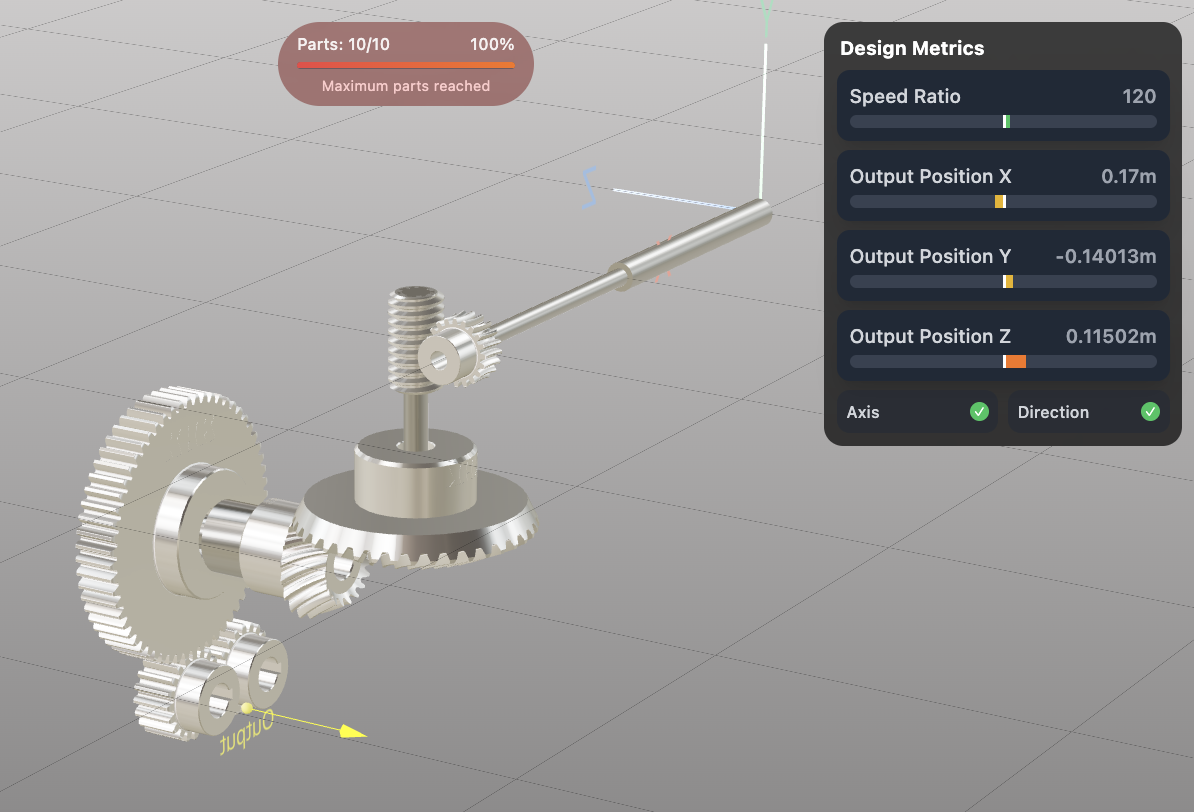}
  \caption{A feasible solution found by a participant (P6) with the Copilot Mode.}
  \label{fig:solution}
\end{figure}

We also calculated the number of participants who were successful in finding a feasible solution, i.e., a solution that met all the requirement metrics, with each mode. Three participants were successful with the Explore Mode while only one participant (P6) was successful with the Copilot Mode. Note that P6 was not able to find a feasible solution with the Explore Mode. While finding a suitable solution is the ultimate goal, the study focused more on understanding participants' experiences navigating a non-trivial problem; thus, the relatively low completion rates in the limited time were somewhat expected and not a primary concern for evaluating the interfaces themselves. Figure~\ref{fig:solution} shows the sole successful solution found with the Copilot Mode.

\begin{figure}[h!]
\includegraphics[width=\columnwidth]{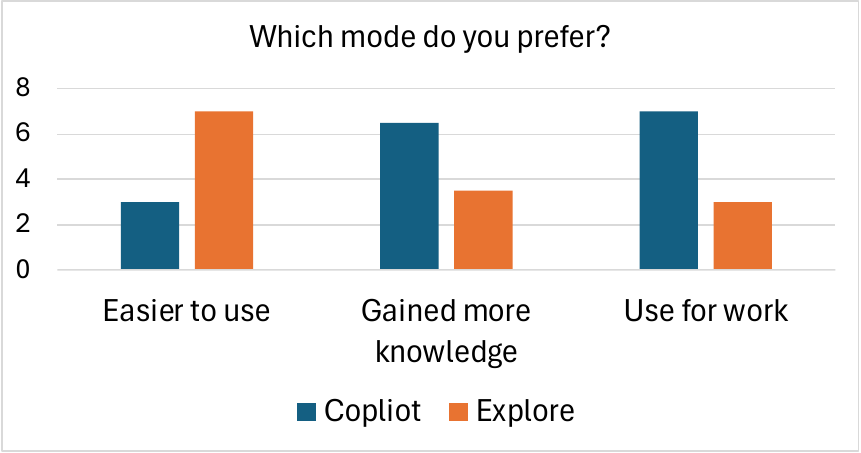}
  \caption{Participants' preferences on the mode that 1) is easier to use, 2) helped gain more knowledge, and 3) they would use for everyday work. }
  \label{fig:pref}
\end{figure}

Figure~\ref{fig:pref} summarizes the responses to the interview questions on comparing the two modes. Seven out of 10 participants found the Explore Mode to be easier to use. On the other hand, 6.5 out of 10 participants thought they gained more knowledge with the Copilot Mode while seven out of 10 participants would prefer to use the Copilot Mode for their daily work. These results show the inherent trade-offs in the capabilities offered by the two modes.

\subsection{Qualitative Findings}

Here we present seven interesting qualitative observations made during the study:

\paragraph{Rapid solution generation}

The Explore Mode offers the capability of quickly producing multiple design alternatives (at around 1 second per design) to be considered by the user. The learning curve was minimal as participants only needed to input the target requirement values and compare different solutions generated, much akin to a typical online search process.

\begin{participantquote}{Participant 6}
``[I like] the fact that it gave me a solution without having to click a load of buttons, which is helpful, [...] to see and compare not just with the gear [designs], but, [with] clear objectives and, obviously [they are met by] a lot of them.''
\end{participantquote}

\begin{participantquote}{Participant 7}
``I think it was useful [...] that I can just specify the design objectives and then the number of iterations and I just click run. And I really liked how it had the bar for each criteria that it was meeting and that I can go ahead and rank them depending on what I'm hoping to optimize. [...] I think the interface is very intuitive. The fact that it's showing each of the gear assemblies, I'm able to look through them very quickly.
\end{participantquote}

On the other hand, it took participants much longer to learn how to use the Copilot Mode, and this led to most participants completing only one or two design solutions within the limited amount of time given. This naturally led to more participants successfully finding feasible solutions in the Explore Mode, as shown in Figure~\ref{fig:success}.

\begin{participantquote}{Participant 7}
``There's more variability in this interface. [...] It took more effort navigating and it's definitely harder to meet the objectives while trusting what's recommended with the positioning and also the parts.''
\end{participantquote}

\paragraph{Domain knowledge integration}

In both modes, we observed how participants' existing knowledge and intuition about gear train designs came into play while interacting with the UI.

In the Explore Mode, when ranking solutions, participants prioritized certain requirements over another, e.g., speed, because they noticed that the position targets can be easily met by making further adjustments on the design.

\begin{participantquote}{Participant 8}
``[Position error] is a potentially a small fix because you can just extrude this [shaft].''
\end{participantquote}

In the Copilot Mode, the user could make component selections based on other design considerations such as torque and structural requirements, supply chain, and manufacturing.

\begin{participantquote}{Participant 3}
``I don't think the model kind of currently considers like, the internal stresses or like the structural integrity of the shafts and all that kind of stuff, which I think might be a big consideration in these sort of designs.

The ability to [...] put in external intuition and design considerations while still having [...] some guidance on what's the most appropriate in terms of the metrics the model understands. I look at the model as like someone who thinks only about these metrics and I'm the person who has to think about all the other metrics.''
\end{participantquote}

One experienced gear designer (P1) offered concerns on how the recommended designs may not meet other important design criteria and highlighted an opportunity to blend in such knowledge as part of the design workflow.

\begin{participantquote}{Participant 1}
``Ultimately the position of these gears with respect to each other is dependent on the total tolerance stack based on the manufacturing process I'm using. So I may need to adjust some of that clearance and that's not communicated to me here yet.

[Regarding] the ability to transmit the torque I need, I have zero confidence right now only because I don't have that information provided I don't know what the load rating of any of these gears are.''
\end{participantquote}

Another experienced gear designer (P2) suggested that they could first think of an initial \emph{draft} design based on their experience and use it as a reference in the Copilot Mode.

\begin{participantquote}{Participant 2}
``If I was to do it again, I would probably sketch something out myself first to figure out what I wanted and [...] have an idea of where I needed to go, because when I started building it, I had no idea where I needed to be.''
\end{participantquote}

\paragraph{Control and transparency}

Five out of 10 participants explicitly stated that with the Copilot Mode, felt they were more in control.

\begin{participantquote}{Participant 4}
``I think the copilot at least gave me the illusion of more control. Even though it's recommending parts, I do still have that ability to select. [...] I personally would feel a bit of a preference for the Copilot though, just because I'd feel like I have some control in my destiny.''
\end{participantquote}

\begin{participantquote}{Participant 5}
``So you know, naturally with the design copilot you have a lot more control. And I think as an experienced designer [...] you want to probably exert that level of control.''
\end{participantquote}

Related to the previous insight, this affordance of more control could allow the user to blend in their own knowledge with the recommendations made by the AI model to make optimal decisions.

However, a few participants also desired for more explanations and reasons for the recommendations made by the AI model.

\begin{participantquote}{Participant 2}
``I don't trust anyone to ever just make a decision for me. I want to know, what is the thing we're dealing with? So like, what is the size of diameter, what is the number of teeth? But being able to see a little bit more of the information that is necessary for designing the tool up front and also knowing why I would choose a 1.5 module versus a 2.5 might be helpful just to understand where I'm going.''
\end{participantquote}

\begin{participantquote}{Participant 6}
``So [the Copilot Mode] might not give you as much knowledge as you think. [...] I think maybe giving someone a bit more information about why that gear was selected and what it's doing might give someone a bit more information and knowledge.''
\end{participantquote}

\paragraph{Design exploration via problem redefinition}

None of the participants were able to find a feasible solution in the Explore Mode in their first iteration of generations.
To obtain different solutions in the next iteration, most participants changed the target requirement values rather than simply increasing the number of generations. In general, participants observed patterns in current iteration designs and adjusted the target values accordingly, e.g.,

\begin{participantquote}{Participant 3}
``So I think the biggest issue was X kept going too far out. [...] Y is sort of in the range. Z is sort of in the range. So maybe X is the problem. [...]. Everything seems to overshoot X a bit. So I'm going to try and bring X down a bit and generate.''
\end{participantquote}

This strategy of adjusting the problem definition resembles problem reformulation occurring in a design process, which has been identified as an essential ingredient for design creativity \cite{dorst2001creativity}.


In contrast, none of the participants attempted to change the input requirement values in the Copilot Mode. This could be largely due to the limited time and participants were allocating all of their time in generating one or two solutions. Nonetheless, these observations may imply that the Copilot Mode might bias the user to focus on solution generation rather than problem definition, potentially hindering the creativity promoted by the latter process.
\paragraph{Over-reliance on AI}

In the Copilot Mode, some participants seemed to fully trust the AI model and simply chose the top-pick recommended by the model.

\begin{participantquote}{Participant 9}
``I don't really know too much about gears, so I am going to trust the AI and I'm going to select the top pick.''
\end{participantquote}

\begin{participantquote}{Participant 8}
``I guess you have no idea what you're doing. It kind of just helps you move forward step by step in [gear] design. So there wasn't much thinking in terms of [which exact] gear do I need to find? [...]
So I guess it [reduces] that cognitive load in your brain.''
\end{participantquote}

The quote by P8 was particularly interesting because it highlighted the fact that the AI model could be a double-edged sword. While it could reduce the cognitive load on the designer, someone without much expertise could blindly follow the AI recommendations without much critical thinking.

One experienced gear train designer (P2) was seen over-relying on the AI recommendations by simply choosing the top pick. At one point, the model attributed equal probability values for all the placement options and P2 become indecisive. It was interesting to observe that P2 did not try to use his own intuition but rather became frustrated by the AI's recommendations.

\begin{participantquote}{Participant 2}
``[...] because this is quite an important decision, which one you're going to pick there, isn't it? I'm surprised it doesn't tell you which [placement] is the best way to choose.''
\end{participantquote}




\paragraph{Opportunity for a hybrid workflow}

Finally, a couple of participants pointed out if there could be a Hybrid mode where a promising solution is first identified with the Explore Mode and that solution is further improved in the Copilot Mode. These observations allude to the potential UI design where both approaches are offered in combination to take advantage of their respective strengths.

\begin{participantquote}{Participant 3}
``So one design that I like [from the Explore Mode] and then [...] take that design into the copilot would also be really nice. [Then] backtrack a bit and try to fix a part of it.''
\end{participantquote}

\begin{participantquote}{Participant 5}
P5: ``Is it possible to go into the Explore Mode and then bring it into this design Copilot Mode?''
\end{participantquote}

\section{Discussion}

Based on the user study results and insights observed, we highlight the benefits and limitations of each interaction mode developed for the UI. We then discuss potential improvements we could make as future work. The discussion is summarized in Table~\ref{tab:summary}.

\begin{table*}[t]
    \centering
    \caption{Summary of the benefits, limitations, and future improvements for each interaction mode.}
    \begin{tabularx}{0.98\textwidth}{X p{5cm}  p{5cm} p{5cm}}
        \toprule
        \textbf{Mode} & \textbf{Benefits} & \textbf{Limitations} & \textbf{Future Improvements} \\
        \midrule
        \textbf{Explore} &
        \begin{minipage}[t]{\linewidth}
            \begin{itemize}[leftmargin=*, itemsep=0pt, topsep=0pt, partopsep=0pt, parsep=0pt]
                \item Easy to use with a minimal learning curve
                \item Supports design space exploration via rapid solution generation and problem redefinitions
            \end{itemize}
        \end{minipage} &
        \begin{minipage}[t]{\linewidth}
            \begin{itemize}[leftmargin=*, itemsep=0pt, topsep=0pt, partopsep=0pt, parsep=0pt]
                \item Lack of perceived control
            \end{itemize}
        \end{minipage} &
        \begin{minipage}[t]{\linewidth}
            \begin{itemize}[leftmargin=*, itemsep=0pt, topsep=0pt, partopsep=0pt, parsep=0pt]
                \item Create a hybrid workflow: Export designs to the Copilot Mode
                \item Allow the user to specify preferences for different requirements
            \end{itemize}
        \end{minipage} \\
         \midrule
        \textbf{Copilot} &
        \begin{minipage}[t]{\linewidth}
            \begin{itemize}[leftmargin=*, itemsep=0pt, topsep=0pt, partopsep=0pt, parsep=0pt]
                \item User feels in-control and easy to integrate domain knowledge
                \item Automates non-creative operations and reduces cognitive load
            \end{itemize}
        \end{minipage} &
        \begin{minipage}[t]{\linewidth}
            \begin{itemize}[leftmargin=*, itemsep=0pt, topsep=0pt, partopsep=0pt, parsep=0pt]
                \item Lack of transparency and explanation on AI recommendations
                \item User could fixate on AI recommendations and solution generation
            \end{itemize}
        \end{minipage} &
        \begin{minipage}[t]{\linewidth}
            \begin{itemize}[leftmargin=*, itemsep=0pt, topsep=0pt, partopsep=0pt, parsep=0pt]
                \item Provide additional information and preview of the solution path along with recommendations
                \item Identify and nudge the user when they are fixated
            \end{itemize}
        \end{minipage} \\
        \bottomrule
    \end{tabularx}
    \label{tab:summary}
\end{table*}

\subsection{Explore}

The biggest advantage of the Explore Mode is that the participants found it very easy to use (Figure~\ref{fig:pref}) and a minimal learning curve. All participants were able to input the target requirement values and quickly obtain a number of design solutions without any issues. It was also easy to redefine the problem by playing around with different target requirement values and generate a new set of design solutions. In practice, these features would allow users to quickly go through multiple iterations of design solutions and increase the chance of discovering an optimal solution (Figure~\ref{fig:success}).

However, the majority of participants expressed that they felt a lack of control compared to the Copilot Mode (Figure~\ref{fig:pref}). Overall, one could argue that the Explore Mode resembles a generic search process rather than a typical design process. Because almost all of the creative aspects of design have been off-loaded to the AI model, there is little room for the user to integrate their domain knowledge or preferences and may not feel much ownership in the final design.

One immediate improvement we could make is to create a hybrid workflow, as suggested by a couple of participants, where the user could first explore different solutions in the Explore Mode and take the most promising solutions to the Copilot Mode to further improve the solutions. This would allow the user to integrate their domain knowledge and exert their control during the second step. In addition, we could align the AI model to respect different user preferences. For example, \citet{cheong2025simft} has shown that GearFormer can be fine-tuned to prioritize addressing specific design requirements. This would support the user to specify preferences on addressing a particular requirement over another, e.g., speed over output position, giving them more control on the solutions being generated.

\subsection{Copilot}

As expressed by the majority of the participants (Figure~\ref{fig:pref}), the Copilot Mode gave them a better sense of control. This allows the user to blend in their domain knowledge and intuition while making the component and placement choices while creating the gear train assembly. For example, many participants did not always go with the top choice recommended by the AI model but went with one of the other choices because they felt that the top choice was too expensive (P6) or would lead to unnecessary complexity (P2). Another interesting comment made by P8 was that the AI model reduces the cognitive load, e.g., by automating non-creative operations such as placing the gears in the scene to make sure they are oriented and meshed appropriately.

On the other hand, participants desired more information and explanation on why certain AI recommendations are made. For example, the top choice recommended might be a bevel gear but no information is provided why that was a good choice. Another negative aspect is that while the Copilot Mode automates many operations including the decision-making, it led to participants going into an \emph{autopilot} mode where they were blindly accepting the top choice recommended by the AI model. Also, none of the participants went back to redefine their problem if the current design creation did not work out.

To increase the transparency behind the recommendations made, the UI could provide additional information that is associated with the recommended choice with respect to the current state of the design.
For example, whenever a bevel gear is recommended, the UI could communicate ``Bevel gears can be used to change the motion axis. Currently, your motion axis is not the same as the target motion axis'' to provide to the user some justification behind the recommended choice. In addition, one participant suggested it would be beneficial to see the preview of a complete assembly design if the particular current choice was made. The AI model already has the capability to provide this information so it would be easy to add this feature to the UI. Finally, it would be useful to have a monitoring module that could identify the user getting fixated on certain procedural aspect and nudge them toward design exploration.

\section{Conclusion}
In this paper, we explored how user interfaces can leverage the unique capabilities of transformer-based generative models to enhance human-AI collaboration in mechanical assembly design tasks. By presenting two distinct interaction modes—Explore Mode and Copilot Mode—we demonstrated how transformer properties such as probabilistic sampling and autoregressive prediction could be effectively utilized to augment human creativity, control, and exploration within complex design spaces.

The findings from our user study indicated that while Explore Mode provided an accessible and intuitive way to rapidly generate and evaluate multiple design solutions, Copilot Mode afforded greater user control, enabling participants to incorporate domain-specific knowledge and iteratively refine designs. Despite Explore Mode being perceived as easier-to-use, participants preferred Copilot Mode for professional use, highlighting the importance of interactive guidance, transparency, and meaningful control over AI recommendations.

Moreover, the insights suggested a promising direction for hybrid workflows, combining the strengths of both exploration-driven and refinement-driven interfaces. Future research should further investigate such integrated workflows, providing richer explanations for model recommendations, improving transparency to prevent over-reliance on AI, and allowing seamless transitions between exploratory and targeted design phases. Ultimately, thoughtfully designed user interfaces that align closely with transformer model capabilities hold significant promise in enhancing engineering design processes and empowering users to achieve creative, high-quality outcomes.

\bibliographystyle{ACM-Reference-Format}
\bibliography{main}










\end{document}